\documentclass{elsart}
\usepackage{graphicx}
\usepackage{dcolumn}
\usepackage{syntonly}   % checking for proper syntax,
\usepackage{amstext}
\usepackage{amsbsy}
\usepackage{amsopn}
\usepackage{amsfonts}   % for different math-fonts
\usepackage{upref}      % certain type of cross-reference numbers
\usepackage{stmaryrd}  % keine Ahnung wozu ich das package brauchte
\usepackage{bm}
\begin{document}
\begin{frontmatter}

\title{The QCD equation of state near $T_c$ within a quasi-particle model}

\author[FZR,TU]{M. Bluhm},
\author[FZR]{B. K\"ampfer\corauthref{cor}}
\corauth[cor]{Corresponding author.}
\ead{kaempfer@fz-rossendorf.de},
\author[TU]{G. Soff}

\address[FZR]{Institut f\"ur Kern- und Hadronenphysik,
Forschungszentrum Rossendorf,\\
PF 510119, 01314 Dresden, Germany}
\address[TU]{Institut f\"ur Theoretische Physik, TU Dresden, 01062 Dresden, Germany}

\begin{abstract}
We present a description of the equation of state of
strongly interacting matter within a quasi-particle model.
The model is adjusted to lattice QCD data near the deconfinement temperature $T_c$.
%at non-vanishing baryon density. 
We compare in detail
the excess pressure at non-vanishing chemical potential 
and its expansion coefficients with
two-flavor lattice QCD calculations and outline prospects of
the extrapolation to large baryon density.
\end{abstract}

\begin{keyword} 
Equation of state, Strongly interacting matter, Lattice QCD
\PACS 12.38.Mh
\end{keyword}
\end{frontmatter}

Due to the recent progress of first principle lattice QCD calculations, the equation of state (EoS) of strongly interacting
matter is now at our disposal in some region of temperature $T$
and chemical potential $\mu$. Either the overlap improving
multi-parameter re-weighting technique \cite{FK} or the Taylor
expansion or hybrids of them \cite{Allton1,Allton2} deliver the
pressure, entropy density, quark density, susceptibilities etc.
The knowledge of these quantities is of primary importance for a
hydrodynamical description of relativistic heavy-ion collisions,
the confinement transition in the early universe and possible
quark cores in compact neutron stars.
Knowing the phase boundary \cite{Allton1} and the end point
of the first-order deconfinement transition \cite{Fodor} in the
region of non-vanishing chemical potential is particularly interesting for
the envisaged CBM project at
the future accelerator facility FAIR at Darmstadt \cite{CBM}. 
In the planned experiments a systematic investigation
of phenomena of maximum baryon density reachable in heavy-ion
collisions will be attempted.

Apart from lattice QCD calculations as purely numerical technique to obtain the
EoS, also analytical approaches have been invented to understand the basic features.
We mention dimensional reduction, resummed HTL scheme,
$\Phi$ functional approach, Polyakov loop model etc.\ (cf.\ \cite{Pei1}
for a recent survey). A controlled chain of approximations from full
QCD to analytical expressions without
adjustable parameters describing the lattice data would be of desire. 
Success has been achieved
for $T > 2.5 T_c$ \cite{Bla1} at vanishing baryon density. 
In contrast, the range $T \ge T_c$, in particular
close to $T_c$, is covered by phenomenological models \cite{Pes1,Lev1} 
with parameters adjusted to lattice QCD data at $\mu = 0$.
It is the aim of the present paper to compare in detail the quasi-particle model
\cite{Pes1} with the recent lattice QCD data \cite{Allton2}
in the region around $T_c$ by extending the focus on finite baryon density.
For the first time, we present a quasi-particle description of
the expansion coefficients of the excess pressure for the quark-gluon fluid.
Only in such a way an adequate and direct comparison with the lattice QCD results
\cite{Allton2,Ejiri} is possible.

One way to decompose the EoS is writing for the pressure \cite{Allton1,Allton2}
\vskip -6mm
\begin{equation}
p(T, \mu) = p(T, \mu=0) + \Delta p(T,\mu),
\quad
\frac{\Delta
p(T,\mu)}{T^4} = \sum_{i=2}^\infty c_i \left( \frac{\mu}{T} \right)^i.
\label{e:exp}
\end{equation}
\vskip -6mm
$p(T,\mu=0)$ was subject of previous lattice QCD calculations
(cf.\ \cite{Kar2} for the two-flavor case), 
while $\Delta p(T,\mu)$ became accessible only recently
\cite{Allton2,Fodor}. $\Delta p(T,\mu)$ is easier calculable, therefore,
lattice QCD calculations focus on this quantity, instead of focussing on $p(T,\mu)$.
In contrast, our model covers $p(T,\mu=0)$ and $\Delta p(T,\mu)$
on equal footing, therefore, we have $p(T,\mu)$ at our disposal.

The quasi-particle model of light quarks ($q$) and gluons ($g$)
is based on the expression for the pressure
\vskip -12mm
\begin{equation}
\label{e:pres}
p = \sum_{a = q,g} p_a - B(T, \mu), \quad
p_a = \frac{d_a}{6 \pi^2} \int dk \frac{k^4}{E_a(k)}
\left( (f_a^+ (k) + f_a^-(k) \right)
\end{equation}
\vskip -6mm
where $B(T,\mu)$ ensures thermodynamic self-consistency \cite{Pes1},
$s =\partial p / \partial T$, $n_q = \partial p / \partial \mu$,
together with the stationarity condition
$\delta p / \delta m_a^2 = 0$ \cite{Gor1}. The $k$ integrals here and below
run from $0$ to $\infty$.
Explicitly, the entropy density reads $s = \sum_{a = q, g} s_a$ 
with\footnote{
In massless $\varphi^4$ theory such a structure of the entropy density emerges by
resumming the super-daisy diagrams in tadpole topology \cite{Pes4},
and \cite{Pes5} argues that such an ansatz is also valid for
QCD. \cite{Bla1} point to more complex structures, but we
find (\ref{e:pres}, \ref{e:ent}, \ref{e:den}) flexible enough to accommodate the lattice data.
Finite width effects are studied in \cite{Pes33}.
In the $\Phi$ functional approach the following chain of approximations
leads to the given ansatz \cite{Bluhm1}:
(i) two-loop approximation for the $\Phi$ functional;
(ii) neglect longitudinal gluon modes and the plasmino branch,
both being exponentially damped;
(iii) restore gauge invariance and ultra-violet finiteness by arming
the self-energies with HTL resummed expressions;
(iv) neglect imaginary parts in self-energies and Landau damping
and approximate suitably the self-energies in the thermodynamically
relevant region $k \sim T, \mu$. The pressure follows by an integration.}
\vskip -6mm
\begin{equation}
\label{e:ent}
s_a = \frac{d_a}{2\pi^2T}
\int dk k^2 \left( \frac{\left( \frac{4}{3}k^2 + m_a^2 \right)}{E_a (k)}
(f_a^+(k) + f_a^-(k))
- \mu (f_a^+ (k) - f_a^-(k)) \right)
\end{equation}
\vskip -6mm
and the net quark number density is
\begin{equation}
\label{e:den}
n_q = \frac{d_q}{2\pi^2} \int dk k^2 (f_q^+(k) - f_q^-(k))
\end{equation}
\vskip -6mm
with degeneracies $d_q = 12$ and $d_g = 8$
as for free partons and distribution functions
$f_a^{\pm}(k) = (\exp( [E_a(k) \mp \mu]/ T) +S)^{-1}$
with $S = + 1$ ($- 1$) for quarks (gluons). The
chemical potential is $\mu$ for light quarks, while for gluons it is zero.

The quasi-particle dispersion relation is approximated by the asymptotic
mass shell expression near the light cone 
\begin{equation}
E_a^2(k) = k^2 + m_a^2,
\quad
m_a^2(T, \mu) = \Pi_a(k; T, \mu) + (x_a T)^2.
\end{equation}
\vskip -6mm
The essential part is the self-energy $\Pi_a$;
the last term accounts for the masses used in the lattice calculation \cite{Allton2},
i.e., $x_q = 0.4$ and $x_g = 0$.  
First direct measurements of the dispersion relation have been reported
in \cite{Petrezky}. It should be noticed, however, that for the EoS the excitations
at momenta $k \sim T$ matter, for which more accurate measurements are needed.
As suitable parametrization of $\Pi_a$, we employ here the HTL self-energies
with given explicit $T$ and $\mu$ dependencies as in \cite{Pes1}.
The crucial point is to replace the running coupling in $\Pi_a$
by an effective coupling, $G^2(T,\mu)$.\footnote{
As shown in \cite{Pes9}, it is the introduced 
$G^2(T,\mu)$ which allows to describe lattice QCD data near $T_c$, while the use of the
pure 1-loop or 2-loop perturbative coupling together with a more complete description
of the plasmon term and Landau damping restricts the approach to $T \ge 2.5 T_c$.}
In doing so, non-perturbative effects are thought to be
accommodated in this effective coupling. This assumption needs detailed
tests which are presented below.
Note that Eqs.~(\ref{e:pres} - \ref{e:den}) themselves are highly
non-perturbative expressions. Expanding them in powers
of the coupling strength one recovers the first perturbative terms.

The first expansion coefficients 
in Eq.~(\ref{e:exp}) follow from (\ref{e:pres}) as
$c_i = \frac{T^{i-4}}{i!} \frac{\partial^i p}{\partial  \mu^i}\vert_{\mu = 0}$:
\vskip -6mm
\begin{eqnarray}
&& c_2  = \frac{3 N_f}{\pi^2 T^3} \int dk k^2
\frac{e^{\omega}}{(e^{\omega} + 1)^2}, \label{eq.c2} \\
&& c_4 = \frac{N_f}{4 \pi^2 T^3} \int dk k^2
\frac{e^{\omega}}{(e^{\omega} + 1)^4}
\left( e^{2 \omega} - 4 e^{\omega} + 1
- \frac{A_2}{\omega} (e^{2 \omega} - 1) \right), \label{eq.c4} \\
&& c_6 = \frac{3 N_f}{385 \pi^2 T^3} \int dk k^2 
\frac{e^{\omega}}{(e^{\omega} + 1)^6}
\left\{ e^{4 \omega} - 26 e^{3 \omega} + 66 e^{2 \omega} - 26 e^{\omega} + 1  
\right. \label{eq.c6} \\
&& - \frac{10}{3} \frac{A_2}{\omega}
\left( e^{4 \omega} - 10 e^{3 \omega} + 10 e^{\omega} - 1 \right) 
%\nonumber \\ && 
+ \frac{4}{3} \frac{A_2^2}{\omega^2}
\left( e^{4 \omega} - 2 e^{3 \omega} - 6 e^{2 \omega} - 2 e^{\omega} +1 
\right) \nonumber \\
&& \left. + \left( \frac{5}{3} \frac{A_2^2}{\omega^3} - 10 \frac{T^2 A_4}{\omega} \right)
\left( e^{4 \omega} + 2 e^{3 \omega} - 2 e^{\omega} -1 \right) \right\}, \nonumber
\end{eqnarray}
where $\omega = (k^2 + \frac13 T^2 G^2\vert_{\mu = 0} )^{1/2} / T$,
$A_2 = (G^2 / \pi^2 + \frac12 T^2  \partial^2 G^2 / \partial \mu^2) \vert_{\mu = 0}$,
$A_4 = (\frac{1}{\pi^2} \partial^2 G^2 /\partial \mu^2 +
\frac{T^2}{12} \partial^4 G^2 / \partial \mu^4 )\vert_{\mu = 0}$.
(We have not displayed the terms $\propto x_q$ stemming from the lattice masses;
in the calculations presented below, however, these terms are included
to make the model as analog as possible to the lattice performance.)  
$c_j$ with odd $j$ vanish.
In deriving these equations we have used the flow equation \cite{Pes1}
\begin{equation} \label{eq.flow}
a_\mu \frac{\partial G^2}{\partial \mu} +
a_T \frac{\partial G^2}{\partial T} = a_{\mu T},
\end{equation}
\vskip -6mm
where the lengthy coefficients $a_{\mu, T, \mu T} (T, \mu)$ \cite{Bluhm1}
obey $a_T (T, \mu=0) = 0$ and $a_{\mu T} (T, \mu=0) = 0$. This flow equation
follows from a thermodynamic consistency condition.
The meaning of Eq.~(\ref{eq.flow}) is to map $G^2$, given on some
curve $T(\mu)$, e.g., on $T(\mu = 0)$, into
the $\mu$ plane to get $G^2(T,\mu)$ which is needed to calculate
$p$, $s$, $n$ from Eqs.~(\ref{e:pres} - \ref{e:den}) at non-vanishing
values of $\mu$.
The terms needed in Eqs.~(\ref{eq.c4}, \ref{eq.c6}) follow from the flow equation 
and its derivatives yielding
\vskip -6mm
\begin{eqnarray}
&&\partial^2 G^2 / \partial \mu^2 \vert_{\mu = 0} =
\frac{1}{a_\mu} \left. \left( \frac{\partial a_{\mu T}}{\partial \mu}
- \frac{\partial a_T}{\partial \mu} \frac{\partial G^2}{\partial T}
\right) \right\vert_{\mu = 0},  \label{eq.g2}\\
&&\partial^4 G^2 / \partial \mu^4 \vert_{\mu = 0} =  
\frac{1}{a_\mu} \left( \frac{\partial^3 a_{\mu T}}{\partial \mu^3}
- \frac{\partial^3 a_T}{\partial \mu^3} \frac{\partial G^2}{\partial T}
- 3 \frac{\partial^2 a_\mu}{\partial \mu^2} \frac{\partial^2 G^2}{\partial \mu^2}
\right. \label{eq.g4}\\
&& \hspace*{12mm} - \left. \left. \frac{3}{a_\mu} \frac{\partial a_T}{\partial \mu} \left[
\frac{\partial^2 a_{\mu T}}{\partial \mu \, \partial T}
- \frac{\partial^2 a_T}{\partial \mu \, \partial T} \frac{\partial G^2}{\partial T}
-\frac{\partial a_T}{\partial \mu} \frac{\partial^2 G^2}{\partial T^2} 
% \right. \right. \nonumber \\ && \hspace*{12mm} \left. \left. 
- \frac{\partial a_\mu}{\partial T} 
\frac{\partial^2 G^2}{\partial \mu^2} \right] \right) \right\vert_{\mu = 0}. \nonumber
\end{eqnarray} 
\vskip -3mm

We adjust $G^2(T)$ through Eq.~(\ref{eq.c2}) to $c_2 (T)$ from \cite{Allton2}
for $N_f = 2$. We find as convenient parametrization
\begin{equation}
\label{e:G}
G^2(T) = \left\{
\begin{array}{l}
G^2_{\rm 2-loop} (T), \quad T \ge T_c,
\\[3mm]
G^2_{\rm 2-loop}(T_c) + b (1- T / T_c), \quad T < T_c,
\end{array}
\right.
\end{equation}
\vskip -3mm
where $G^2_{\rm 2-loop}$ is the relevant part of the 2-loop coupling
\begin{equation}
\label{eq.G2}
G^2_{\rm 2-loop}(T) = \frac{16 \pi^2}{\beta_0 \log \xi^2}
\left[ 1 - \frac{2 \beta_1}{\beta_0^2} \frac{\log (\log \xi^2)}{\log \xi^2} \right]
\end{equation}
\vskip -3mm
with
$\beta_0 = (11 N_c - 2 N_f) / 3$,
$\beta_1 = (34 N_c^2 -13 N_f N_c + 3 N_f /N_c)/6$, %$\beta_2 = ...$,
and the argument $\xi = \lambda (T - T_s)/T_c$.
$T_s$ acts as regulator at $T_c$, and $\lambda$ sets the scale. 
The parameters for $N_c = 3$ are
$\lambda = 12$, $T_s = 0.87 T_c$, 
and $b = 426.1$.
Fig.~1 exhibits the comparison of $\Delta p$ and $n$ 
calculated via Eqs.~(\ref{e:pres}, \ref{e:den}) (dashed curves)
or by using the expansion coefficients Eqs.~(\ref{eq.c4}, \ref{eq.c6})
(solid curves) with the lattice QCD data \cite{Allton2} based on the 
coefficients $c_{2,4}$ (symbols).
One observes an astonishingly good description of the data, even slightly
below $T_c$, where the resonance gas model \cite{Kar3} 
is appropriate.\footnote{
Some reasoning why the model may be applicable
also slightly below $T_c$ can be found in \cite{Bormio}.}  
Interesting is the deviation of the full model from the results
based on the truncated expansion in a small interval around $T_c$.
It should be noted that conceptionally different models
\cite{Thaler} reproduce the lattice data for $\Delta p$ and $n$
fairly well above $T_c$.
Since for small values of $\mu$ the higher order coefficients
$c_4$ and in particular $c_6$ are less important for $\Delta p$ and $n$,
a more stringent test of the model is accomplished by a direct comparison
of the individual expansion coefficients $c_i$ with the corresponding 
lattice QCD results.
%The expansion coefficients are directly related to the quark number susceptibility
%via $\chi = \partial^2 (p/T^4) / \partial (\mu/T)^2 
%= 2 c_2(T) + 12 c_4 (T) (\mu/T)^2+ 30 c_6(T) (\mu/T)^4$.
%$c_{4,6} $ enter the derivatives of this susceptibility
%and serve as such as sensitive control quantities. 
 
\begin{figure}[t]
\hspace*{-1mm}
\includegraphics[width=5.5cm,angle=90]{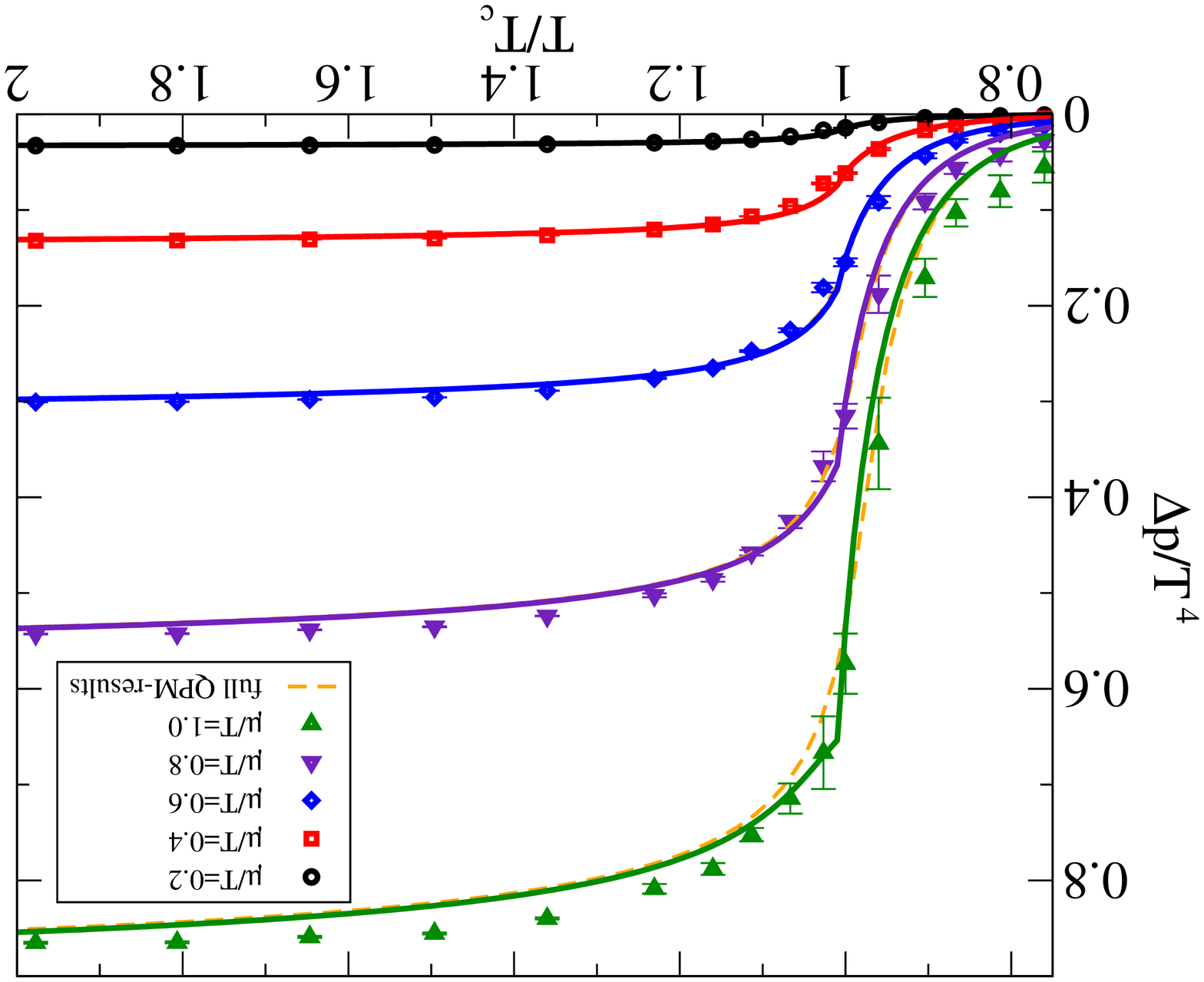}
\hspace*{69mm}
\includegraphics[width=5.5cm,angle=90]{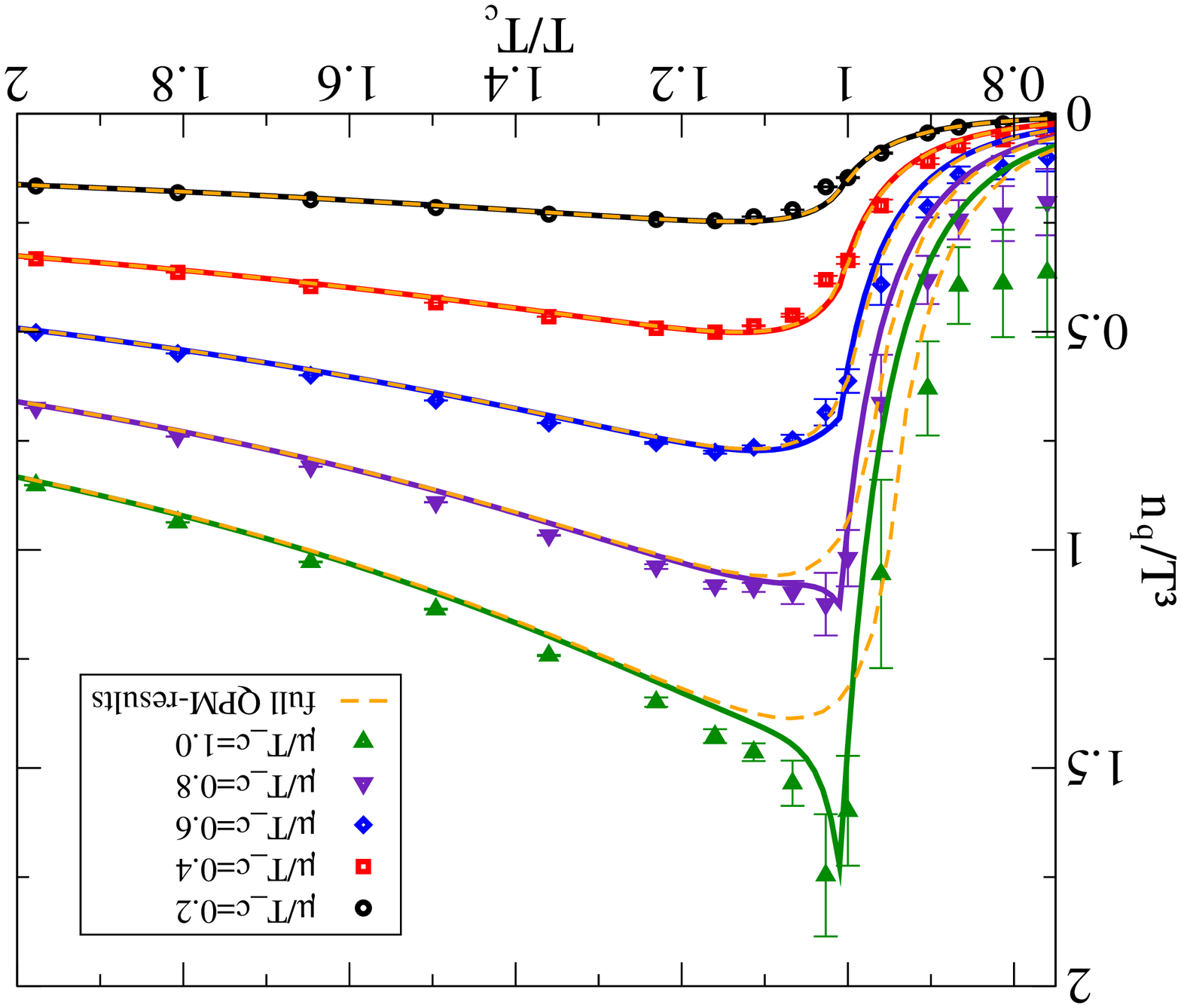}
\caption{
\label{fig:eos}
Comparison of the quasi-particle model with lattice QCD results \cite{Allton2} 
for the excess pressure (left panel, for constant $\mu / T$) 
and net quark number density (right panel, for constant $\mu / T_c$ ).
As for the lattice QCD data (symbols)
the quasi-particle model results (solid curves) are based on the the expansion
coefficients $c_{2,4}$. For comparison, the full quasi-particle model
results (dashed curves) are exhibited.}
\end{figure}

Straightforward evaluation of Eqs.~(\ref{eq.c2} - \ref{eq.c6}) 
delivers the results exhibited in Fig.~2.
Since $G^2(T)$ was already adjusted to $c_2(T)$ the agreement is good.
It should be emphasized that all coefficients $c_i (T)$ 
are determined by $G^2(T)$. That means
%depend on the one function $G^2(T)$ and its derivatives.
the same $G^2(T)$ describes also the features of $c_4$ and $c_6$.
Particularly interesting are the peak of $c_4$ (left panel of Fig.~2)
and the double-peak of $c_6 / c_4$ (right panel of Fig.~2)
or $c_6$ (not exhibited) at $T_c$. Numerically, these pronounced structures stem from
the change of the curvature behavior of $G^2(T)$ at $T_c$
which determines the terms $\partial^2 G^2 / \partial \mu^2\vert_{\mu=0}$ and
$\partial^4 G^2 / \partial \mu^4\vert_{\mu=0}$ via 
Eqs.~(\ref{eq.flow}, \ref{eq.g2}, \ref{eq.g4}).
Neglecting these terms would
completely alter the shape of $c_{4,6}$. Similar to \cite{Allton2},
we interpret the peak in $c_4$ as indicator of some
critical behavior, while the pressure itself is smoothly but rapidly varying
at $T_c$.

\begin{figure}[t]
\hspace*{-3mm}
\includegraphics[width=5.5cm,angle=90]{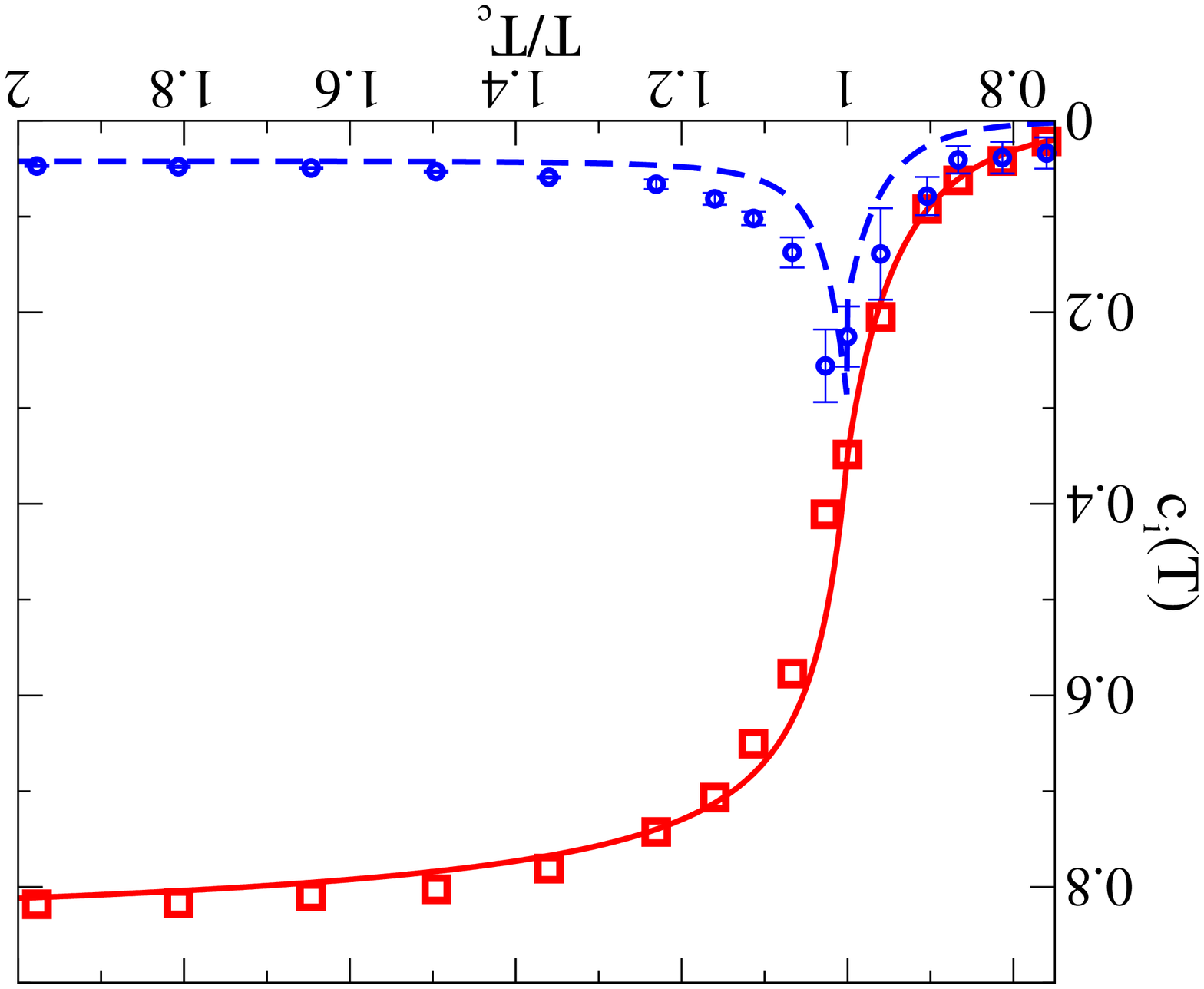}
\hspace*{69mm}
\includegraphics[width=5.5cm,angle=90]{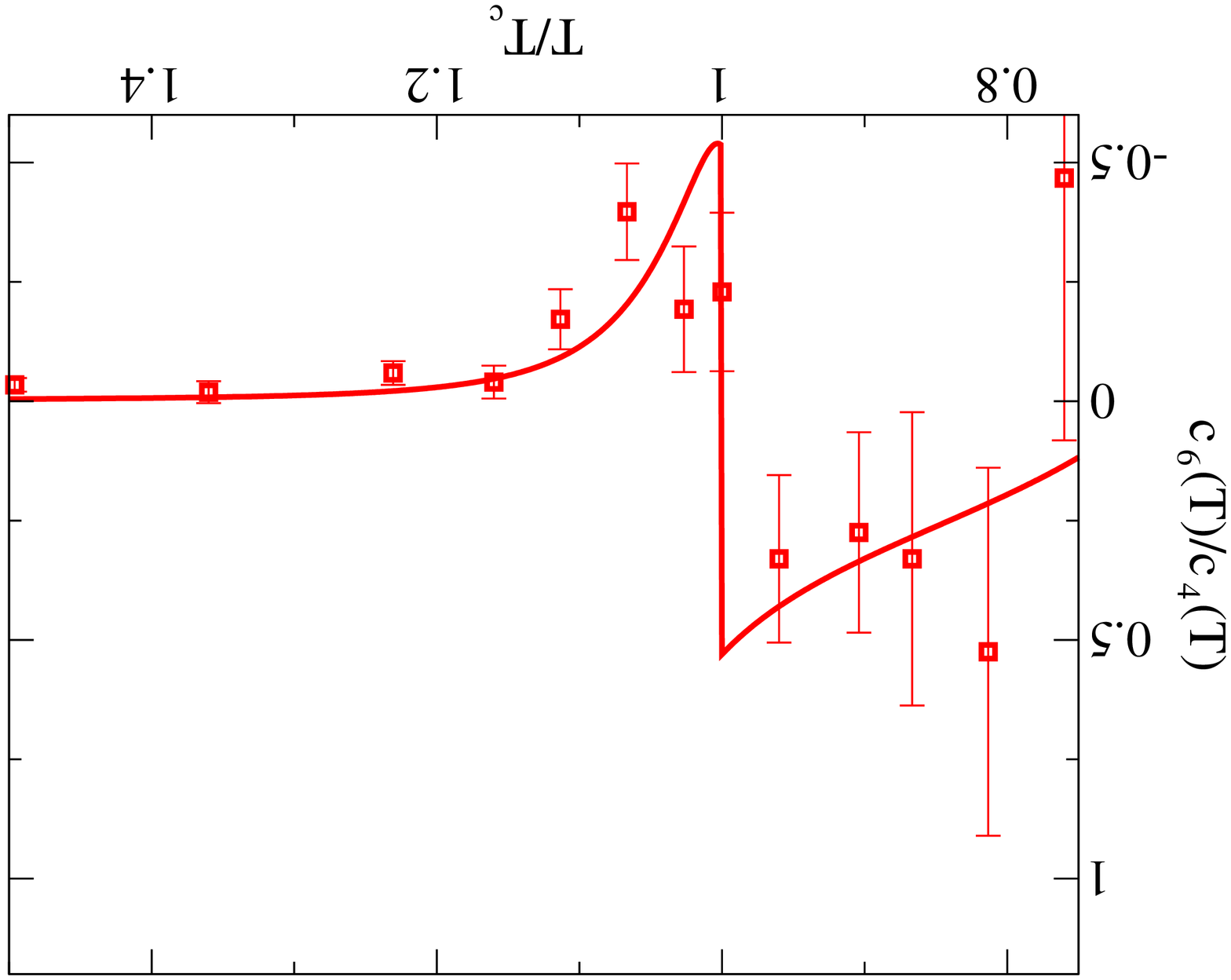}
\put(-44,122){$c_2$}
\put(-44,39){$c_4$}
\caption{
The expansion coefficients $c_{2,4}$ (left panel, data from \cite{Allton2})
and the ratio $c_6/c_4$ (right panel, data from \cite{Ejiri})
as a function of the temperature. 
\label{fig:c24}}
\end{figure}

In summary we present a quasi-particle model which describes the recent
lattice QCD data for non-vanishing chemical potential remarkably well.
Besides the excess pressure $\Delta p(T, \mu)$ and density $n$ above and even slightly
below $T_c$ at small values of the chemical potential, 
the individual expansion coefficients agree well with the data
and turn out to depend on each other. In particular, also $p(T, \mu=0)$
follows, once $G^2(T)$ is adjusted. We find a small
deviation from the data \cite{Kar2} which might be attributed to differences 
in calculating $p(T, \mu=0)$ and the coefficients of $\Delta p(T,\mu)$ on the lattice.

Having tested these details of the quasi-particle model, we can directly
apply the found parametrization and calculate the total pressure at arbitrary
baryon densities, while lattice QCD calculations are yet constraint
to small baryon densities. Such applications are of interest for the
CBM project at FAIR and for studying hot proto-neutron stars and cool
neutron stars with quark cores and will be reported elsewhere.
Another application to cosmic confinement dynamics is reported
in \cite{SQM2004}. These applications need a controlled chiral extrapolation
which must base on improved lattice QCD data. 

Inspiring discussions with F. Karsch, A. Peshier, and K. Redlich
are gratefully acknowledged.
The work is supported by BMBF 06DR121, GSI and EU-I3HP.

\end{document}